\documentclass[prl,aps,twocolumn,superscriptaddress,preprintnumbers,floatfix,longbibliography,nofootinbib]{revtex4-1}
\usepackage{amsmath,amssymb,amsfonts,bm,graphicx,epsf,colordvi,bbm,pifont,siunitx,enumitem,xcolor,multirow}
\usepackage[caption=false]{subfig}
\pdfoutput=1
\usepackage[hyperfootnotes=false]{hyperref}
\hypersetup{
 colorlinks=true,
 citecolor=blue,
 linkcolor=blue,
 urlcolor=blue}
\usepackage[utf8]{inputenc}

\begin{document} 
\title{Extracting the Top-Quark Width from Non-Resonant Production}
\author{Christian Herwig}
\email{herwig@sas.upenn.edu}
\affiliation{Department of Physics and Astronomy, 209 S 33rd St, Philadelphia, PA 19104, USA}

\author{Tom\'a\v{s} Je\v{z}o}
\email{tomas.jezo@physik.uzh.ch}
\affiliation{Physics Institute, Universit\"at Z\"urich, Z\"urich, Switzerland}

\author{Benjamin Nachman}
\email{bpnachman@lbl.gov}
\affiliation{Physics Division, Lawrence Berkeley National Laboratory, Berkeley, CA 94720, USA}

\preprint{ZU-TH 11/19}

\newcommand\sss{\mathchoice%
{\displaystyle}%
{\scriptstyle}%
{\scriptscriptstyle}%
{\scriptscriptstyle}%
}

\begin{abstract}
In the context of the Standard Model (SM) of particle physics, the relationship between the top-quark mass and width ($\Gamma_t$) has been precisely calculated.  However, the uncertainty from current direct measurements of the width is nearly 50\%.  A new approach for directly measuring the top-quark width using events away from the resonance peak is presented.  By using an orthogonal dataset to traditional top-quark width extractions, this new method may enable significant improvements in the experimental sensitivity in a method combination.  Recasting a recent ATLAS differential cross section measurement, we find $\Gamma_t=1.28\pm0.30$~GeV~($1.33\pm0.29$~GeV expected), providing the most precise direct measurement of the width.  
\end{abstract}

\maketitle

\section{Introduction}

Even though the top quark was discovered over 20 years ago~\cite{Abe:1995hr,D0:1995jca} and its mass has been measured with a sub-percent precision~\cite{ATLAS:2014wva}, direct measurements of its width $\Gamma_t$ have an uncertainty of 50\% or worse~\cite{Aaltonen:2013kna,Aaboud:2017uqq,CMS-PAS-TOP-16-019}.  Indirect measurements of $\Gamma_t$ using single top-quark production are more precise, but also require additional modeling assumptions~\cite{Abazov:2012vd,Khachatryan:2014nda}.   In the context of the Standard Model (SM), the relationship between the top-quark mass and $\Gamma_t$ is known at next-to-next-to-leading order (NNLO) in the strong coupling constant with an uncertainty of a few percent~\cite{CZARNECKI1999520,PhysRevD.60.114015,PhysRevLett.93.062001}.  Improving the precision of $\Gamma_t$ can therefore provide a stringent test of the SM.

The current methods for measuring $\Gamma_t$ perform template fits to invariant mass spectra that peak near the top-quark mass.  Due to their cleaner event signatures, the lepton+jets and dilepton decays from $t\bar{t}$ production are used for the fits and the most sensitive observable is $m_{lb}$: the invariant mass of a charged lepton and a jet originating from a $b$ quark ($b$ jet).  While the location of the $m_{lb}$ peak is linearly sensitive to the top-quark mass, the measured width around the peak depends sub-linearly on $\Gamma_t$.  This is because the measured width is dominated by the unmeasured neutrino momentum and the fragmentation of the $b$ quark, not by $\Gamma_t$.

We propose a new method\footnote{Similar ideas were recently discussed in Ref.~\cite{Liebler:2015ipp,Baskakov:2018huw,Baskakov:2017jhb}.  The key difference with respect to this analysis are that these studies (I) were performed at parton-level only, (II) do not propose a physical observable for measuring the cross-section, and (III) do not extract any results with LHC data (and in the case of Ref.~\cite{Liebler:2015ipp}, are for $e^+e^-$).  A related idea using $b$-jet charge asymmetries in $pp$ collisions was proposed in Ref.~\cite{Giardino:2017hva}.  While a promising proposal, this latter study does not yet include reconstruction effects.} for measuring $\Gamma_t$ that is linearly sensitive to $\Gamma_t$.  The idea is motivated by recent proposals to measure the Higgs boson width from non-resonant production~\cite{Kauer:2012hd,Caola:2013yja,Campbell:2013una,Campbell:2013wga}, which has interesting beyond the Standard Model (BSM) sensitivity~\cite{Englert:2014aca,Cacciapaglia:2014rla,Azatov:2014jga,Ghezzi:2014qpa,Buschmann:2014sia,Gainer:2014hha,Englert:2014ffa,Goncalves:2017iub} and has lead to measurements by the CMS~\cite{Khachatryan:2014iha,Khachatryan:2016ctc,Khachatryan:2015mma,Sirunyan:2017exp} and ATLAS~\cite{Aaboud:2018puo} collaborations.  Similarly, we propose to study the $WWbb$ cross section far from the $t\bar{t}$ resonance peak, using a method that can be linearly sensitive to $\Gamma_t$.  Furthermore, this approach may be sensitive to BSM physics that only affects the non-resonant part of the spectrum.  Such modifications could be due to interference effects from complex phases or from new decay channels that are kinematically inaccessible at the resonance peak.

\section{Existing Calculations and Measurements}
\label{sec:status}

Calculations of the top-quark width at next-to-leading order (NLO) in the strong coupling constant were first performed in Refs.~\cite{JEZABEK19891,Czarnecki:1990kv,PhysRevD.43.3759}.  
More recently, the NNLO computation of the total width has been completed \cite{CZARNECKI1999520,PhysRevD.60.114015,PhysRevLett.93.062001}, followed by fully differential calculations of the decay rate \cite{Gao:2012ja,Brucherseifer:2013iv}.
At leading order (LO), the dependence of the width on the top-quark mass is given by 
\begin{equation}
\Gamma_t^\text{LO} = \dfrac{G_F m_t^3}{8\sqrt{2}\pi}\left(1-\frac{m_W^2}{m_t^2}\right)^2 \left(1+2\frac{m_W^2}{m_t^2}\right),
\end{equation}
assuming $|V_{tb}|=1$ and neglecting the $b$-quark mass~\cite{JEZABEK19891}.
For a top-quark mass of 172.5 GeV the predicted width is 1.322~GeV, including NNLO QCD, finite $b$-quark and $W$ masses, and NLO electroweak corrections \cite{Gao:2012ja}.

The width of the top quark has been measured at the Tevatron and the Large Hadron Collider (LHC) using several approaches.
Direct measurements, based on the reconstructed top-quark mass distribution in events with top-quark pairs were made by the CDF~\cite{Aaltonen:2013kna}, ATLAS~\cite{Aaboud:2017uqq}, and CMS~\cite{CMS-PAS-TOP-16-019} Collaborations, with Ref.~\cite{Aaboud:2017uqq} obtaining the most precise value of 
$\Gamma_t = 1.76 \pm 0.33$ (stat.) $^{+0.79}_{-0.68}$ (syst.) using this method. The D0~\cite{Abazov:2012vd} and CMS~\cite{Khachatryan:2014nda} Collaborations have each determined the width indirectly, via a combination of the $t$-channel single-top cross section and measurement of the branching fraction ratio $\mathcal{BR}(t\to Wb)/\mathcal{BR}(t\to Wq)$.  The most precise estimate from Ref.~\cite{Khachatryan:2014nda} finds
$\Gamma_t = 1.36 \pm 0.02$ (stat.) $^{+0.14}_{-0.11}$ (syst.), with the restrictive assumption that $\mathcal{BR}(t\to Wq)=1$.

\section{Sensitivity of the ATLAS Measurement}
\label{sec:atlas}

Recently, the ATLAS Collaboration reported a differential cross section measurement of events with two charged leptons ($\ell=e,\mu$) and two $b$-jets, in an observable sensitive to both top-quark pair ($t\bar t$) and single top-quark ($tWb$) production~\cite{Aaboud:2018bir}.
The measurement probed the interference between $t\bar{t}$ and $tWb$ by comparing the data with state-of-the-art interference models~\cite{Frixione:2008yi,Hollik:2012rc,Demartin:2016axk,Jezo:2016ujg}.
The target observable was the minimax pairing of lepton-jet invariant masses $m_{b\ell}$,
\begin{equation}
m_{b\ell}^\text{minimax} = \min\{ %
\max(m_{b_1\ell_1},m_{b_2\ell_2}),%
\max(m_{b_1\ell_2},m_{b_2\ell_1})\},
\end{equation}
where the labeling of leptons and $b$-tagged jets is arbitrary.
For values of $m_{b\ell}^\text{minimax} > \sqrt{m_t^2-m_W^2}$, the top-quark pair process at LO enters only through off-shell effects and $tWb$ contributions become important.
In this high-$m_{b\ell}^\text{minimax}$ region, the NLO calculation of $bb\ell^+\nu_{\ell}l^-\overline{\nu}_{l}$ including interference effects~\cite{Bevilacqua:2010qb, Denner:2010jp, Denner:2012yc,Heinrich:2013qaa, Frederix:2013gra, Cascioli:2013wga} and parton showering~\cite{Jezo:2016ujg}, provides an excellent description of the data.

The advent of such calculations enable these data to constrain other properties of the top quark.
Specifically, modifications to the top-quark width impact the $m_{b\ell}^\text{minimax}$ spectrum. 
The origin of this dependence is twofold.

First, the cross section of events with $m_{b\ell}^\text{minimax}$ considerably greater than $\sqrt{m_t^2-m_W^2}$ has a contribution from top-quark pair production diagrams, where at least one of the top (anti-)quarks is produced far off-shell.
The width impacts the size of this contribution directly through the top-quark lineshape, which can be described as a Breit-Wigner distribution:
\begin{equation}
\frac{d\sigma}{dm^2_{Wb}} \sim \frac{1}{(m_{Wb}^2-m_t^2)^2 + m_t^2\Gamma_t^2}.
\end{equation}
Integrating over both top-quark resonances, the fraction of off-shell events is found to be linearly dependent on the width~\cite{Kauer:2001sp}.

Second, the ``tail'' cross-section is also enhanced by $tWb$ diagrams containing only one top-(anti-)quark propagator.
While this is a smaller overall contribution than that of top-quark pairs, the $W$ boson and $b$ quark not originating from a top quark often satisfy $m_{Wb}>m_t$, so that a comparatively large fraction of such events have $m_{b\ell}^\text{minimax}>\sqrt{m_t^2-m_W^2}$.
For this reason, width variations affect the relative importance of these two classes of diagrams and thus the shape itself of the $m_{b\ell}^\text{minimax}$ observable.

In Ref.~\cite{Aaboud:2018bir}, the fractional contribution of $WWbb$ events to 15 bins of $m_{b\ell}^\text{minimax}$ was reported, including many measurements with $m_{b\ell}^\text{minimax} > \sqrt{m_t^2-m_W^2}$.
Despite the measurement not considering a width uncertainty, the unfolded result would only be impacted through migrations in the response matrix, and the effect is thus expected to be negligible.  For comparison, the uncertainty due to unfolding with different interference schemes is $<5\%$ in most bins, despite the predictions leading to raw differences of 50\% or more for large values of $m_{b\ell}^\text{minimax}$.

\section{Signal models and event selection}
\label{sec:signal}

The primary calculation used to model the $W^+W^-b\bar{b}$ signal is the \texttt{b\_bbar\_4l}~\cite{Jezo:2016ujg} generator implemented in \texttt{POWHEG\,BOX\,RES}~\cite{Jezo:2015aia}.
It includes NLO QCD corrections and matching with the parton shower (PS) based on the \textsc{Powheg} method~\cite{Nason:2004rx, Frixione:2007vw}.
The process is described in terms of exact matrix elements for $pp \to \ell^+\nu_{\sss\ell}\, l^-\bar{\nu}_{\sss l} \, b \,\bar b $, dominated by top-pair topologies with leptonic $W$-boson decays, with massive $b$ quarks provided by the \textsc{OpenLoops} program~\cite{Cascioli:2011va}.
The \texttt{b\_bbar\_4l} generator combines for the first time: consistent NLO+PS treatment of top-quark resonances, including quantum corrections to top-quark propagators and off-shell top-quark decay chains; exact spin correlations at NLO, interference between NLO radiation from top-quark production and decays, full NLO accuracy in $t \bar{t}$ production and decays; NLO accuracy in $b$-quark kinematics; access to phase-space regions with unresolved $b$ quarks and/or jet vetoes.

The nominal event sample was obtained using the \texttt{NNPDF30\_nlo\_as\_0118} Parton Distribution Function (PDF) set and the input top-quark mass value $m_t = 172.5$~GeV.
Additional samples were generated with a range of top-quark widths $\Gamma_t\in\{0.66,1,\Gamma_t^{\text{SM}},1.66,2\}$~GeV (with $m_t = 172.5$~GeV and\footnote{The value of $\Gamma_t^{\text{SM}}$ is the NLO top-quark width calculated by \texttt{b\_bbar\_4l} from all the other input values.} $\Gamma_t^{\text{SM}} = 1.3328$ GeV) to enable the extraction of the best-fit value from data, with a range of top-quark mass values $m_t\in\{171.5,172.5,173.5\}$~GeV (with $\Gamma_t = \Gamma_t^{\text{SM}})$ in order to estimate the uncertainty due to the top-quark mass, and a range of $\alpha_S$ values in the PDF $\alpha_S\in\{0.115,0.118,0.121\}$ (with $m_t = 172.5$~GeV and $\Gamma_t = \Gamma_t^{\text{SM}})$ to explore the uncertainty due to variation of scale of the shower evolution.
The central renormalization and factorization scales are set to the geometric average of transverse masses of the top quark and anti-top quark, and the value of \texttt{hdamp} is always set equal to the input value of the top-quark mass.
The samples include all possible combinations of different family final state leptons (corresponding to the \texttt{channel 7} setting).
Events also feature additional weights due to standard 7-point scale variation and due to PDF variation.
Up to three hardest emissions are kept with the \texttt{allrad 1} setting, one from the production process and one from each of the top-quark resonances, and matching to \textsc{Pythia\,8.2} makes use of both the \textsc{PowhegHooks} and \textsc{PowhegHooksBB4L}~\cite{Ravasio:2018lzi} vetoes, and A14 set of tuned parameters~\cite{ATL-PHYS-PUB-2014-021}.
In the samples with $\alpha_S\in\{0.115,0.121\}$ the 
\texttt{SpaceShower:alphaSValue} parameter of shower evolution in \textsc{Pythia\,8.2} is set correspondingly.

In addition, a LO calculation of the $W^+W^-b\bar{b}$ process is examined, calculated by 
\texttt{Madgraph5\_aMC@NLO}~2.6.4 with up to 2 extra jets, matched to a parton shower implemented in \textsc{Pythia\,8.240}. 
This sample of events was simulated using the \texttt{NNPDF23\_nlo\_as\_0118} PDF set, the A14 set of tuned parameters, $m_t = 172.5$~GeV, and $\Gamma_t\in\{0.66,1,1.33,1.66,2\}$~GeV.  Alternative samples were produced with $\alpha_S$ varied as described above, as well as with alternative top-quark mass hypotheses $m_t\in\{170,175\}$~GeV.

Event samples are analyzed and compared to data using the selection criteria of Ref.~\cite{Aaboud:2018bir} as implemented in the \texttt{Rivet} toolkit~\cite{Buckley:2010ar}.
Briefly, leptons and jets are reconstructed at particle-level with selections based upon the acceptance of the ATLAS detector.  
Leptons are dressed with nearby photons and are required to have transverse momentum $p_T>28$~GeV and pseudorapidity $|\eta|<2.47$ (2.5) for electrons (muons). Jets are reconstructed with the anti-$k_T$ algorithm using a radius parameter of $R=0.4$~\cite{Cacciari:2008gp,Cacciari:2005hq,Cacciari:2011ma} and considered in the analysis only if $p_T>25$~GeV and $|\eta|<2.5$. They are $b$-tagged if a $B$-hadron with $p_T>5$~GeV is found within the jet cone.
Events are selected which have two leptons, two $b$-tagged jets, with same-flavor lepton events vetoed if the dilepton mass $m_{\ell\ell}<10$~GeV or satisfies $|m_{\ell\ell}-m_Z|<15$~GeV.

The \texttt{b\_bbar\_4l} simulation produces events with different-flavor leptons and must be corrected to account for same-flavor contributions.  The $ee$ and $\mu\mu$ contribution is obtained by re-weighting the generated $e\mu$ events which satisfy same-flavor $m_{\ell\ell}$ requirements.  Good closure of this technique is found using the LO 
\texttt{Madgraph5\_aMC@NLO} simulation, which includes all leptonic decays of the $W$ boson.  Further, the contribution of di-$\tau$ events (with fully leptonic $\tau$ decays) is found to be negligible and is not considered.

\section{Top-quark width extraction}
\label{sec:method}

Using the experimental data of Ref.~\cite{Aaboud:2018bir} and the signal models described above, the top-quark width is extracted by minimizing the following $\chi^2$ statistic:
\begin{equation}
\label{eq:chi2}
\chi^2 = \sum_{i,j} (d_i-m_i) \cdot V_{ij}^{-1} \cdot (d_j-m_j),
\end{equation}
where $d_i$ is the measured, normalized,
differential cross section indexed by bins of $m_{b\ell}^\text{minimax}$ and $m_i$ is the corresponding prediction.
The covariance matrix $V_{ij}$ gives the uncertainty on the unfolded data, including bin-to-bin correlations.
The measurements with $m_{b\ell}^\text{minimax} < 160$~GeV are only weakly sensitive to variations in $\Gamma_t$ and are thus excluded from Eq.~\ref{eq:chi2}.

For each systematic uncertainty, the differential cross section is computed separately for a set of test widths $\Gamma_t$.
To interpolate between generated samples, the calculated yields are fit as a function of the top-quark width to obtain a parameterized prediction $m_i = F_i(\Gamma_t)$, individually for each bin $i$.
Choosing the functions $F_i$ to be quadratic in $\Gamma_t$ is found to fit the calculated predictions well for each bin of $m_{b\ell}^\text{minimax}$.
Thus, given the data and choice of signal model, the statistic may be written explicitly as a function of the width $\chi^2=\chi^2(\Gamma_t\mid d,m)$. By minimizing this function with respect to $\Gamma_t$, the best-fit value of the width may be extracted.

\section{Uncertainties}
\label{sec:uncerts}

Uncertainties stemming from the precision of the experimental measurement, from choices in signal modeling, and from the limited number of generated events are each considered.
Pseudo-experiments are used to assess the experimental uncertainty, where pseudo-data are drawn from a multivariate gaussian distribution with mean and covariance matrix given by $d_i$ and $V_{ij}$. 
For each pseudo-experiment a random dataset $d_i^\text{pseudo}$ is drawn from this distribution and a new value of $\Gamma_t$ is extracted by minimizing $\chi^2(\Gamma_t \mid d_i^\text{pseudo},m_i)$.  
The experimental uncertainty is calculated as the $1\sigma$ range of extracted widths from the pseudo-experiment distribution.

Theoretical uncertainties are assessed on the \texttt{b\_bbar\_4l} signal model by generating event samples with alternative input parameters.  The nominal simulated sample with alternative weight sets is used to estimate the uncertainty due to the choice of PDF as well as renormalization and factorization scales.  The PDF uncertainty is assessed as the standard deviation of widths extracted over the set of 100 eigenvector variations of the \texttt{NNPDF30\_nlo\_as\_0118} PDF set.  The scale uncertainty is the maximum pairwise difference between the widths extracted with the nominal and varied scales.

For top-quark mass and $\alpha_s$ variations, independent samples of events are generated.
To minimize the impact of statistical variations across samples and make optimal use of all generated events, the systematic uncertainty dependence is extracted in a fit, writing
\begin{equation}
m_i(\alpha_s,m_t) = \hat m_i(\alpha_s^\text{SM},m_t^\text{SM}) 
 + \hat a_i(\alpha_s-\alpha_s^\text{SM})
+ \hat b_i(m_t-m_t^\text{SM}).
\end{equation}
Further, the fitted coefficients $\hat a_i$, $\hat b_i$ are constrained to vary quadratically in $m_{b\ell}^\text{minimax}$ to reduce unphysical, statistical fluctuations.
The post-fit yields for $m_t$ and $\alpha_s$ variations are then used to re-weight the nominal $m_{b\ell}^\text{minimax}$ spectra for each value of the top-quark width and to extract the $\chi^2$-minimizing value for each variation.

An uncertainty due to the finite number of simulated events is estimated from an ensemble of pseudo-experiments where the predicted yields for all bins of each value of the top-quark width are varied within their uncertainties.  A width is obtained for each trial to assess the impact on the final extracted parameter.

For the \texttt{MG5\_aMC@NLO} signal model, an identical set of uncertainties are assessed, employing the same estimation methods, with the following modification: the \texttt{NNPDF23\_nlo\_as\_0119} PDF set is used as the nominal value for this sample.  
The top-quark mass uncertainty is assessed using samples with $m_t=170,175$~GeV, interpolating to obtain the same 1~GeV variations as used above.

\begin{figure}[h!]
\centering
\includegraphics[width=0.5\textwidth]{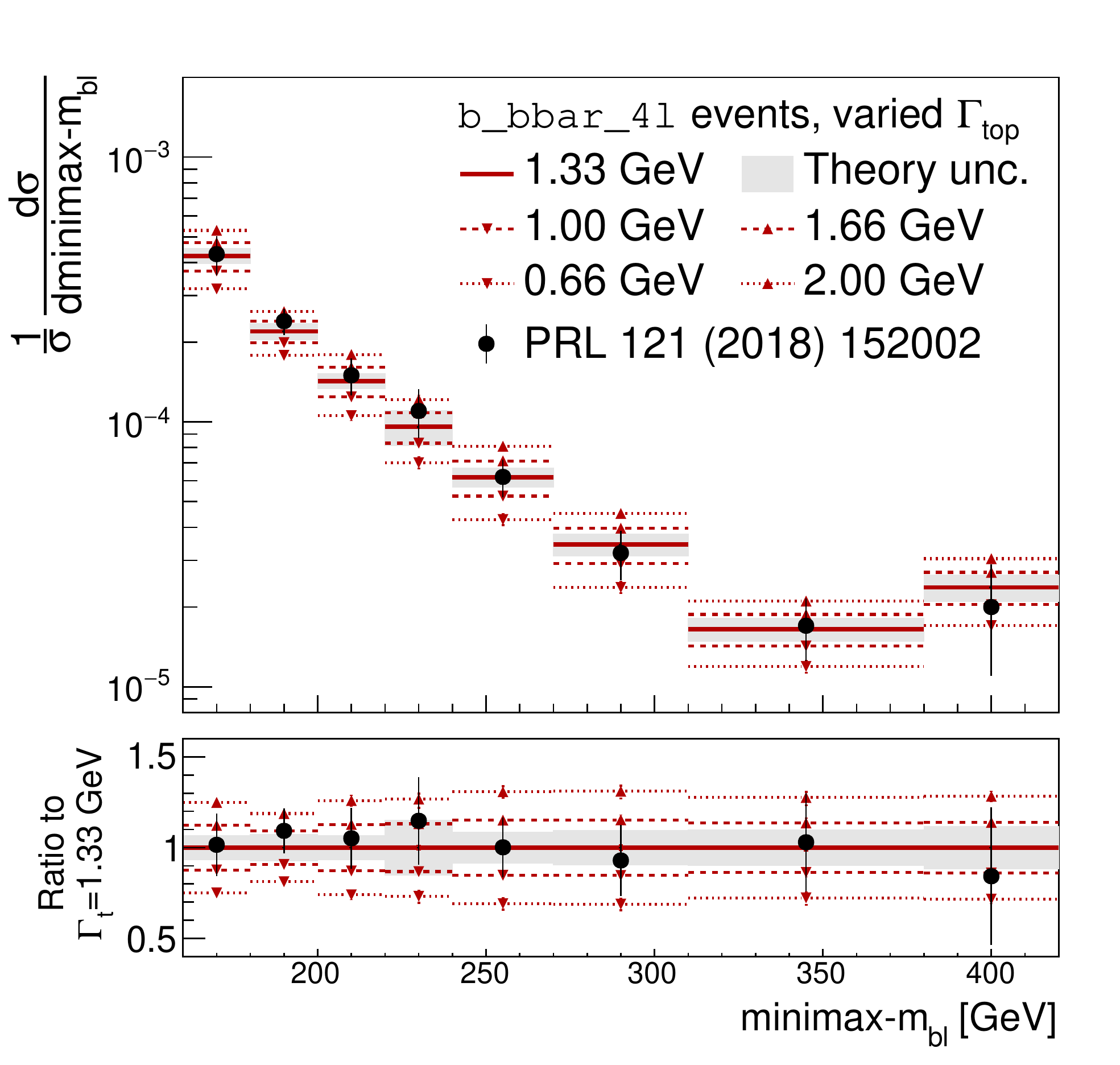}
\caption{%
The $m_{b\ell}^\text{minimax}$ spectrum predicted using \texttt{b\_bbar\_4l} is shown for various values of the top-quark width.  Data from the unfolded ATLAS measurement are included for comparison.  The grey band shows the theoretical uncertainty for the simulated sample corresponding to the predicted SM value of the width.}
\label{fig:mbl}
\end{figure}

The nominal \texttt{b\_bbar\_4l} prediction is compared to ATLAS data in Figure~\ref{fig:mbl}.  Predictions for alternate values of the top-quark width are also shown, as well as the theoretical uncertainty on the nominal estimate.
A summary of the uncertainties on the width extracted using both signal models is presented in Table~\ref{tab:uncerts}.
Changes to the top-quark width are found to produce larger variations in the relative fraction of events in the $m_{b\ell}^\text{minimax}$ tail for samples generated using \texttt{MG5\_aMC@NLO} than \texttt{b\_bbar\_4l}.
As a result, the impact of uncertainties on the extracted width parameter is generally smaller when using the LO simulation, despite the impact on the normalized differential cross section being similar.
This effect leads to a smaller uncertainty due to scale variations, among others, in the LO sample than in the more accurate \texttt{b\_bbar\_4l} calculation.

 \begin{table}[h!]
   \centering
	 \caption{Uncertainty on the top-quark width extracted for data, with individual contributions shown from experimental, theoretical, and statistical sources.
 \label{tab:uncerts}
   }
   \begin{tabular}{ l | l | c | c}
     \multicolumn{2}{ l |}{Uncertainty [GeV]}   & \texttt{b\_bbar\_4l} & \texttt{MG5\_aMC@NLO} \\ \hline
     \multicolumn{2}{ l |}{Experimental}& +0.27/-0.26 & $\pm$0.20 \\ \hline
     \multirow{5}{*}{Theory} & PDF & $\pm$0.06 & $\pm$0.04 \\ \cline{2-4}
      & Scale & $\pm$0.10 & $\pm$0.06 \\ \cline{2-4}
      & $m_t$ & $\pm$0.03 & $\pm$0.03 \\ \cline{2-4}
      & $\alpha_s$ & $\pm$0.06 & $\pm$0.04 \\ \cline{2-4}
      & Combined & $\pm$0.14 & $\pm$0.10 \\ \hline
     \multicolumn{2}{ l |}{Simulation Stats.}& $\pm$0.04 & $\pm$0.04 \\ \hline
     \multicolumn{2}{ l |}{Total}& $\pm$0.30 & $\pm$0.22 \\ \hline
   \end{tabular}
 \end{table}

\section{Results}
\label{sec:results}

\begin{figure}[h!]
\centering
\includegraphics[width=0.5\textwidth]{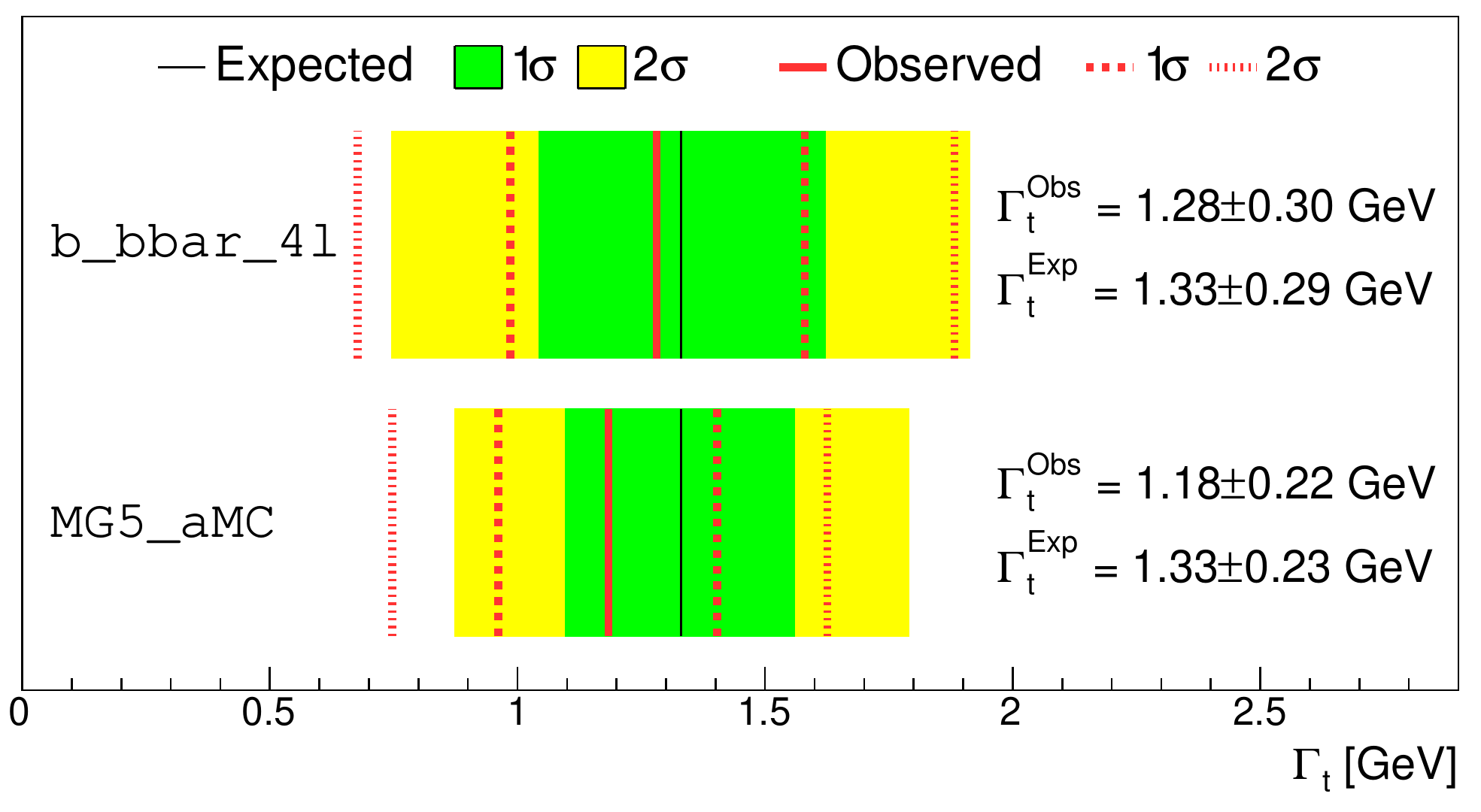}
\caption{%
Observed and expected top-quark widths for the \texttt{b\_bbar\_4l} and \texttt{MG5\_aMC@NLO} signal models.}
\label{fig:result}
\end{figure}

Using the \texttt{b\_bbar\_4l} signal description, a top-quark width of 
$1.28\pm0.30$~GeV is extracted ($1.33\pm0.29$~GeV expected), as shown in Figure~\ref{fig:result}.
A width is also extracted using the leading order \texttt{MG5\_aMC@NLO} simulation, obtaining
$1.18\pm0.22$~GeV ($1.33\pm0.23$~GeV expected).
These measurements are more precise than the previously most precise direct measurement of ($1.76^{+0.86}_{-0.76}$~GeV)~\cite{Aaboud:2017uqq}.

These results can also be interpreted into a limit on the BSM decays of the top-quark through the relation
\begin{equation}
\mathcal{BR}(t\to\text{BSM}) < \frac{\Gamma_\text{ext}^\text{+95\%} - \Gamma^\text{SM}_{t\to bW}}{\Gamma_\text{ext}^\text{+95\%}},
\end{equation}
where $\Gamma^\text{SM}_{t\to bW}$ is the SM partial width for $t\to Wb$ and $\Gamma_\text{ext}^\text{+95\%}$ is the (one-sided) upper limit on the top-quark width at the 95\% confidence level.
The limit is $\mathcal{BR}(t\to\text{BSM})<29\%$ using the \texttt{b\_bbar\_4l} model (30\% expected) and
18\% using the \texttt{MG5\_aMC@NLO} model (26\% expected).  Model-specific BSM constraints are also possible for processes which have a significant effect in the off-shell region defined by high $m_{b\ell}^\text{minimax}$.  For example, a charged Higgs $H^+$ produced via its $btH^+$ coupling and then decaying via $\tau\nu$ would preferentially enhance this region.  However, limits from this measurement are not as strong as direct searches~\cite{Aaboud:2018gjj, Aaboud:2018cwk, Khachatryan:2015qxa} because of the penalties from the $\tau$ decay to $e/\mu$.

\section{Conclusions}
\label{sec:conclusions}

In conclusion, we present a novel method to directly measure the top-quark width and have provided a first estimate using the technique based on a recent measurement of top-quark properties by the ATLAS Collaboration.
The estimate of $1.28\pm0.31$ GeV obtained using the \texttt{POWHEG\,BOX\,RES} calculation is in good agreement with the Standard Model prediction of 1.32 GeV and more precise than existing direct measurements.  Future measurements with more data and in combination with other extractions will be able to provide robust constraints on the top sector of the SM.

\acknowledgments
We  would  like  to  thank Till Eifert, Silvia Ferrario Ravasio, Jay Howarth, Elliot Lipeles, and Stefano Pozzorini for careful reading and suggestions on the manuscript.
We would also like to thank Stefan Prestel for a clarifying conversation about the Pythia treatment of resonances.
B.N. is supported by the DOE under contract DE-AC02-05CH11231.
C.H. is supported by the DOE under contract DE-SC0007901.
The work of T.J. is supported in part by the University of Z\"urich under the contract K-72319-02-01 and in part by the Swiss National Science Foundation under contract BSCGI0-157722.

\appendix

\bibliography{myrefs}

\end{document}